\documentclass[prx,a4paper,superscriptaddress,twocolumn,amsmath,amssymb,floatfix,amstex]{revtex4-1}%

\newcommand{\ud}{\mathrm{d}}

\usepackage{graphicx,graphics,rotating}	% Include figure files
\usepackage{dcolumn}
\usepackage{bm}
\usepackage{hyperref}
\usepackage{color}
 \usepackage{soul}

\begin{document}
%\title{Localization-delocalization transition and multifractality in random smallworld networks}
%\title{Scaling theory of the Anderson transition in random quantum networks:\\ universality and ergodicity}
\title{Kane-Fisher weak link physics in the {clean scratched-XY model}}

\author{G.~Lemari\'e}
\email[Corresponding author: ]{lemarie@irsamc.ups-tlse.fr}
\affiliation{Laboratoire de Physique Th\'eorique, IRSAMC, Universit\'e de Toulouse, CNRS, UPS, France}
\affiliation{Department of Physics, Sapienza University of Rome, P.le A. Moro 2, 00185 Rome, Italy}

\author{I. Maccari}
\affiliation{ISC-CNR and Department of Physics, Sapienza University of Rome, P.le A. Moro 2, 00185 Rome, Italy}

\author{C. Castellani}
\affiliation{Department of Physics, Sapienza University of Rome, P.le A. Moro 2, 00185 Rome, Italy}

\date{\today}% 

\begin{abstract} {The nature of the superfluid-insulator transition in 1D has been much debated recently. In particular, to describe the strong disorder regime characterized by weak link proliferation, a scratched-XY model has been proposed [New J. Phys. \textbf{18}, 045018 (2016)], where the transport is dominated by a single anomalously weak link and is governed by Kane-Fisher weak link physics. In this article, we consider the simplest problem to which the scratched-XY model relates: a single weak link in an otherwise \textit{clean} system, with an intensity $J_W$ which decreases algebraically with the size of the system $J_W\sim L^{-\alpha}$. Using a renormalization group approach and a vortex energy argument, we describe the Kane-Fisher physics in this model and show that it leads to a transition from a transparent regime for $K>K_c$ to a perfect cut for $K<K_c$, with an adjustable $K_c=1/(1-\alpha)$ depending on $\alpha$. 
We check our theoretical predictions with Monte Carlo numerical simulations complemented by finite-size scaling. Our results clarify two important assumptions at the basis of the scratched-XY scenario, the behaviors of the crossover length scale from weak link physics to transparency and of the superfluid stiffness.}  
\end{abstract}
%%%

\maketitle

\section{Introduction}

Understanding the effects of disorder on 1D quantum bosonic systems is a very challenging {issue} \cite{giamarchi2004quantum}. Without interactions, we know that even an infinitesimal degree of disorder leads to Anderson localization \cite{Anderson:PR58}. But what {happens} when interactions compete with disorder is  {much} less clear. At first, one could expect that a collective superfluid state would be immune to {weak} disorder. {Indeed}, theoretical studies in 1D \cite{Giamarchi1987a, Giamarchi1988a} and higher dimensions \cite{Fisher:dirtyboson:PRB89} have shown that the competition between disorder and interaction leads to a superfluid-insulator transition. Understanding this transition is important because it is relevant for many different types of experimental systems, such as Josephson junction arrays \cite{PhysRevLett.119.167701}, spin ladders \cite{PhysRevLett.101.137207} or cold atoms \cite{PhysRevLett.113.095301}.

The nature of the superfluid-insulator transition has been much debated recently, with different scenarios put forward: a weak-disorder regime with a Berezinskii-Kosterlitz-Thouless {(BKT)} transition characterized by a jump of the Luttinger liquid parameter at the universal value $K_c=3/2$ \cite{Giamarchi1987a, Giamarchi1988a, ristivojevic2012phase}; and a new strong disorder regime governed by weak link physics and a non-universal $K_c>3/2$. In this strong disorder regime, a real space renormalization group approach \cite{Altman2004a, Altman2008a, Altman2010a, refael2013strong} and a ``scratched-XY model'' incorporating a Kane-Fisher renormalization of weak links \cite{Yao2016a, pfeffer2018strong} have been proposed to describe the new properties of the superfluid-insulator transition. 

More precisely, according to \cite{Altman2004a, Altman2008a, Altman2010a, refael2013strong}, the regime of strong disorder {induces effectively a power law distribution of weak links}, which can be seen as abnormally weak Josephson couplings between superfluid puddles \cite{vosk2012superfluid}. These weak links, denoted as $J$ then have a power law distribution $P (J) \sim J^{\gamma} $ and {an effective model \cite{Altman2004a, Altman2008a, Altman2010a, refael2013strong} suggests that the inverse of the superfluid density can be written as the average of the inverse weak link couplings $ {\rho_s}^{-1} =\sum_i {J_i}^{-1}/L $ with $L$ the system size. However, because the inverse weak links ${J_i}^{-1}$ do not have a second moment for {$\gamma \le 1$}, the central limit theorem {does not apply} \cite{bouchaud1990anomalous}. On the contrary, the superfluid density may be dominated by the weakest link $J_\text{min}$ over the $L$ weak links. For power law distributed weak links, the weakest link scales as a power law with system size $J_\text{min} \sim L^{-\alpha}$, with $\alpha = 1/(\gamma+1)$. In this case, the superfluid density is predicted to vanish as $\rho_s\sim L^{\gamma/(\gamma+1)}$ for $\gamma<0$, leading to an insulating state \cite{refael2013strong}. This is a very different machanism, based on large disorder fluctuations rather than the proliferation of phase slips {(see \cite{PhysRevB.53.13091})} associated to the BKT transition in the weak disorder regime \cite{giamarchi2004quantum, Giamarchi1987a, Giamarchi1988a, ristivojevic2012phase}.}

However, this argument does not take into account a possible Kane-Fisher renormalization of weak links {(see \cite{Yao2016a})}. Indeed, a striking prediction of Kane and Fisher \cite{Kane1992a, kane1992transmission} is that a weak link in an otherwise clean Luttinger liquid sees its {effective strength $ \mathcal J_W \sim J_W L^{-1/K} $ decrease with system size $L$ so that the system is perfectly transparent $\rho_s \sim L \mathcal J_W \rightarrow 1$ when $K>1$, whereas for $K <1$, the weak link cuts the system in two, $\rho_s \rightarrow 0$.} 
{Kane-Fisher physics has important experimental consequences e.g. for fractional quantum Hall edge states \cite{PhysRevLett.71.4381, PhysRevLett.74.3005, PhysRevLett.79.2526, de1997r} and the delicate crossover it implies has been studied in different 1D quantum systems recently \cite{PhysRevLett.112.106601, vasseur2017healing}}. 
In \cite{Yao2016a}, the authors proposed a scratched-XY model where the transport in a given disordered sample of size $ L $ is dominated by the weakest link {$J_\text{min} \sim L^{-\alpha}$. They then suppose that Kane-Fisher physics applies to this situation, which should make the weakest link $J_\text{min}$ weaker: $\mathcal J_\text{min} \sim L^{-\alpha-1/K}$. Then $\rho_s \sim L^{1-\alpha -1/K}$ and a transition to an insulating phase is possible at $\alpha = 1-1/K_c <1$, thus {even} for $\gamma>0$ contrary to the previous analysis \cite{Altman2004a, Altman2008a, Altman2010a, refael2013strong}, with a critical value of the Luttinger parameter
$K_c = 1/(1-\alpha)$ which can be larger than $3/2$ for $\alpha>1/3$.} 

Despite several numerical studies \cite{pielawa2013numerical, PhysRevLett.109.265303, gerster2016superfluid, doggen2017weak}, there is no consensus today on the strong disorder scenario. In particular, in \cite{doggen2017weak}, {using extensive numerical simulations by the density matrix renormalization group and quantum Monte Carlo approaches}, two different regimes of the BKT superfluid-insulator transition have been observed. At weak disorder, a Giamarchi-Schulz regime is observed where $K_c=3/2$ and the superfluid density and the single particle correlator are self-averaging at criticality. On the contrary, the strong disorder regime is qualitatively different with a proliferation of weak links, $K_c>3/2$ and self similar power-law {critical} distributions for the superfluid density and correlator characterized by {the same} exponent $\gamma$. While this work clearly validates a number of theoretical predictions made previously \cite{Giamarchi1988a, refael2013strong, Yao2016a}, it differs with the strong disorder scenarios {of Refs.~\cite{refael2013strong} and \cite{Yao2016a}} on two important points. The critical values of $\gamma$ have been found significantly larger than $0$ ($\gamma>2.3$), in contradiction with the strong disorder renormalization group approach \cite{refael2013strong}, and the value of $K_c$ is much larger than $1/(1-\alpha)$ predicted by the scratched-XY model \cite{Yao2016a}.

{In order to better understand the origin of these differences, we wanted to consider in detail, the simplest problem to which the scratched-XY model \cite{Yao2016a} relates: a single weak link in an otherwise \textit{clean} system, with an intensity $J_W$ which decreases algebraically with the size of the system $J_W\sim L^{-\alpha}$. The predictions of \cite{Yao2016a} are indeed crucially based on the physics of this model, and in particular (i) on the existence of a characteristic length called ``clutch scale'' which describes the Kane-Fisher crossover physics, and (ii) on the assumption of a ``classical flow'' equation for the superfluid density. In \cite{Yao2016a}, the clutch scale is derived from phenomenological arguments and the validity of the ``classical flow'' approximation used has not been checked. In this article, we shall give {an analytical} derivation of the clutch scale and of the crossover flow for the superfluid density which are {then assessed} by numerical simulations. } 

{To describe the effect of a power-law weak link $J_W\sim L^{-\alpha}$ on a clean Luttinger liquid, we use the analogy between 1D quantum systems and the classical 1+1 XY model, where the additional dimension corresponds to the imaginary time in the quantum problem \cite{giamarchi2004quantum}. The 2D XY model can be understood as an effective model describing the phase fluctuations associated to the 1D quantum case. Since the weak link potential term does not depend on the imaginary time {in this analogy}, the weak link is transposed into a vertical column of weak links.
We treat this problem analytically by a {renormalization group approach and vortex energy arguments}. We also perform numerical simulations by the classical Monte Carlo approach, complemented by finite-size scaling, to check carefully the analytical predictions.} 

Our results confirm the key predictions of \cite{Yao2016a} in the case of a single weak link $J_W\sim L^{-\alpha}$ in a \textit{clean} system, in particular that $ K_c = 1 / (1- \alpha) $. {Importantly}, this allows us to characterize the Kane-Fisher transition in the classical 2DXY model. {Indeed, a necessary condition for Kane-Fisher physics is that the bulk of the system is quasi-ordered, which in the classical 2DXY model requires $K>2$ {due to the Berezinskii-Kosterlitz-Thouless transition that arises at $K=2$ towards a disordered phase} \cite{berezinskii1972destruction, kosterlitz1973ordering, kosterlitz1974critical}. By considering sufficiently large values ​​of $ \alpha >0.5 $, we can work in a regime where the threshold for the Kane-Fisher transition $ K_c> 2 $ and thus observe both the transparent and the cut regimes of Kane-Fisher physics.}

{The paper is organized as follows. In section II, we describe the 2DXY model with a columnar weak link and the analytical and numerical approaches used to describe Kane-Fisher physics in this system. Section III describes the well known case of a constant weak link: we detail our renormalization group predictions for the evolution with system size of the effective weak link strength, and give in particular an analyical expression for the clutch scale assumed in \cite{Yao2016a}. We check these predictions with our numerical Monte Carlo results. In section IV, we describe the evolution with system size of the stiffness, and assess numerically the ``classical flow'' assumed in \cite{Yao2016a}. Section V describes the new case of a power-law weak link whose strength decreases algebraically with system size. We show in particular that a Kane-Fisher transition can be observed in the 2DXY model at $K_c=1/(1-\alpha)$ for $\alpha>0.5$. Section VI gives a complementary vortex energy argument for the Kane-Fisher transtion at $K_c=1/(1-\alpha)$. Section VII discusses the implications of these results on the scratched-XY scenario and concludes.}

\section{The classical 2DXY model with a columnar weak link}

\subsection{Classical 2DXY Model versus 1D quantum bosonic systems}

The classical 2DXY model consists of planar rotors of unit lenth on a two dimensional lattice. The Hamiltonian is given by:
\begin{equation}\label{eq:H2DXY}
 H = -J \sum_{<i,j>} \cos(\theta_i - \theta_j)
\end{equation}
where {$J$ is the coupling constant,} $<i,j>$ denotes nearest neighbors {on a square lattice of spacing set to $a=1$} and $\theta_i$ the angle of the rotor on site $i$ with respect to some (arbitrary) direction in the two dimensional vector 
space of the rotors. 

At low temperature, statistical fluctuations involve only long-wavelength modes {\cite{berezinskii1972destruction, kosterlitz1973ordering, kosterlitz1974critical, benfatto2013berezinskii}}. We can use a continuum approach, which means replacing
the Hamiltonian of the classical 2DXY model
by:
\begin{equation}\label{eq:H2dXYcont}
 \mathcal H = \frac{1}{2} J \int (\boldsymbol\nabla \theta)^2 d\boldsymbol r \; .
\end{equation}

Hence, we can understand the link between the classical 2DXY model and 1D quantum bosonic systems \cite{giamarchi2004quantum}. One can write the partition function of the quantum system as a classical
field path integral:
\begin{equation}
Z=\int D\Psi(x,\tau) D\Psi^*(x,\tau) e^{-S/\hbar} \; ,
\end{equation}
where $\Psi(x,\tau)$ is a complex number field which depends both on $x$ and $\tau$ the immaginary time.
The field $\Psi(x,\tau) = \sqrt{\rho(x,\tau)} e^{i\theta(x,\tau)}$ can be written as a function of the density $\rho$ and the phase $\theta$. 
Usually, the superfluid to insulator transition is driven by phase fluctuations. In a low-energy, long-wavelength description, we can write an effective action \cite{popov2001functional, PhysRevB.53.13091}
which describes the slow variations in the phase of the order parameter:
\begin{equation}\label{eq:LuttS}
 S_\text{eff}[\theta] =\int dx \; d\tau  
 \left[\frac{\rho_s}{2} (\partial_x \theta)^2 +\frac{\kappa}{2} (\partial_\tau \theta)^2  \right] \; .
\end{equation}
Here $\rho_s$ is the superfluid density and 
$\kappa$ compressibility of the 1D quantum system. This action is equivalent to a $1+1$ classical XY model with the immaginary time direction replaced
by the $y$ direction. {The so-called Luttinger parameter $K= \pi \sqrt{\rho_s \kappa}$ corresponds to $K=\pi J/T$ in the classical 2DXY model which controls the algebraic decay of the 
correlation function $\langle \cos(\theta_i -\theta_j) \rangle \sim r_{ij}^{-1/2K}$ where $r_{ij}$ denotes the distance between the two sites $i$ and $j$. In the classical 2DXY model, the Berezinskii-Kosterlitz-Thouless transition arises at the universal value $K=2$.}

\subsection{Columnar weak link}\label{sec:columnarWL}

We are interested in the 2D classical analog of a weak link in 1D quantum bosonic systems. A weak link can be seen as an exponentially weak 
Josephson coupling between two superfluid systems, described by the following term:
\begin{equation}\label{eq:JosCoup}
 H_W = -J_W \cos(\theta_L - \theta_R) \; ,
\end{equation}
where $\theta_{L,R}$ is the phase at the left/right side of the weak coupling. In the mapping from 1D quantum to 1+1 classical systems, an important property is that 
such potentials do not depend on immaginary time $\tau$ \cite{giamarchi2004quantum}. Therefore, the classical analog of the weak Josephson coupling \eqref{eq:JosCoup} is a columnar weak link, 
translation invariant along $y$:
\begin{equation}\label{eq:Hweaklink}
 \mathcal H_W = -J_W \int \text{d}y \cos(\theta_L(y) - \theta_R(y)) \; . 
\end{equation}
In the discrete 2DXY model, this is equivalent to consider a column between say $x=0$ and $x=1$ with $J=J_W$ for all $y$, while $J=1$ otherwise, 
and periodic boundary conditions along the $x$ and $y$ directions.

\subsection{Analytical and numerical methods}

In the following, we will describe the effect of a columnar weak link on the 2DXY model using two approaches. Analytical calculations are performed using on the one hand
a perturbative renormalization group approach {and a self-consistent harmonic approximation}; on the other hand a {vortex energy} argument. 
This is complemented by numerical simulations using a classical
Monte Carlo method \cite{landau2014guide} similar to that used in \cite{maccari2017broadening, maccari2018bkt}{, complemented by a finite-size scaling approach}.

{In the Monte Carlo approach we used, a single Monte Carlo step consists of five Metropolis spin flips of the whole lattice, needed to probe the correct canonical distribution of the system, followed by ten over-relaxation sweeps of all the spins, which help the thermalization leaving unchanged the energy (microcanonical spin sweep). For each temperature we perform {up to $\sim 170 \cdot 10^3$} Monte Carlo steps, and we compute a given quantity  averaging over the last {$160\cdot 10^3$ steps},  discarding  thus the transient regime which occurs in the first {$10^4$} steps {(the Monte Carlo correlation time for the stiffness is less than $20$ steps in the case of the largest system sizes $L=256$ considered)}.} 

{The two observables numerically computed are: \\
(i) the superfluid stiffness $\rho_s$ along the $x$ axis: 
}
\begin{eqnarray}
\label{js}
\rho_{s}&=&\mathcal J_d- \mathcal J_p \; ,\\
\label{jd}
\mathcal J_d&=& \frac{1}{L}\left \langle \sum_{i=1}^{L} J_{i,i+x}\cos(\theta_i - \theta_{i+x}) \right\rangle \; , \\
\label{jp}
\mathcal J_p&=& \frac{\beta}{L} \left \langle \left [\sum_{i=1}^{L} J_{i,i+x} \sin(\theta_i - \theta_{i+x}) \right]^2 \right\rangle \; ,
\end{eqnarray}
{
where $\langle \dots \rangle$ stands for the average over the thermodynamical ensemble (the stiffness $\rho_y$ along the y axis was computed using the previous formula with $x$ replaced by $y$);
(ii)  the correlation function across the weak link:
}
\begin{equation}
C_W= \frac{1}{L}\left \langle \sum_{j=1}^{L} \cos(\theta_{L,j}-\theta_{R,j}) \right\rangle \; .
\end{equation}

\section{Kane-Fisher renormalization of a columnar weak link}\label{sec:constWLcorr}

Kane-Fisher physics \cite{Kane1992a, kane1992transmission} considers the transport through a single impurity/weak link in a 1D quantum system described by the Luttinger liquid theory, 
i.e. by an effective action such as \eqref{eq:LuttS} characterized by the Luttinger parameter $K=\pi \sqrt{\rho_s \kappa}$.
In the non-interacting limit, corresponding to $K=1$, it is well known that an incoming plane wave will
be partially reflected and partially transmitted, with a transmission probability which is a non-trivial number between $0$ and $1$. On the contrary, 
Kane and Fisher showed that for an interacting 1D quantum system at the thermodynamic limit, 
the transmission is either perfect for $K>1$ (i.e. for attractive interactions) or vanishes in the 
repulsive case $K<1$. This physics has important experimental consequences e.g. for fractional quantum Hall edge states \cite{PhysRevLett.71.4381, PhysRevLett.74.3005, PhysRevLett.79.2526, de1997r}.

{The problem we are interested in concerns a power-law weak link whose strength $J_W$ vanishes algebraically with system size. Before describing it, we will first consider the well known case of a constant weak link.} Kane-Fisher physics has now been solved by non-perturbative \cite{PhysRevB.70.075102, PhysRevB.71.155401} or exact \cite{PhysRevB.94.115142} analytical methods, but we will resort here to a more standard perturbative renormalization group approach. We will compare our theoretical predictions with Monte Carlo numerical simulations.

The strategy we followed to describe the effect of a weak link on a column \eqref{eq:Hweaklink}
in the 2DXY model was first (i) to study the relevance of \eqref{eq:Hweaklink} as a perturbation on a decoupled system (corresponding to $J_W=0$){, and (ii) to describe the full crossover by taking into account $J_W$ explicitly. Point (ii) will allow us to give an analytical expression for the clutch scale which describes the crossover physics as confirmed by our numerical results. }

\paragraph*{(i)} Take a decoupled system described by the Hamiltonian \eqref{eq:H2dXYcont} and use the standard identity for Gaussian distributed variables:
\begin{equation}\label{eq:cosexp}
 \langle \cos(\theta_R-\theta_L)\rangle = e^{-  \frac{\langle (\theta_R-\theta_L)^2 \rangle}{2}} \; .
\end{equation}
In the decoupled case, clearly $\langle \theta_R \theta_L \rangle = 0$. {Moreover}, the fluctuations of $\theta_R$ and $\theta_L$ are stronger than in the bulk because
they lie at the boundary. Indeed, the boundary condition $\left.\frac{\partial \theta}{\partial x}\right\vert_{x=0} = 0 $ implies that
the usual Fourier decomposition performed to calculate $\langle \theta \theta \rangle$ has to be reformulated using $\cos(q x)$ instead of plane waves. After some calculations \cite{giamarchi2004quantum, cazalilla2004bosonizing}, one finds that  
\begin{equation}\label{eq:thetatheta}
\langle \theta_R \theta_R \rangle = \langle \theta_L \theta_L \rangle = 2 \langle \theta \theta \rangle_\text{bulk} = \frac{1}{K} \ln L\; ,
\end{equation}
with $K= \pi J/T$. Note that here and in the following, the system size $L$ is dimensionless, measured as a function of a microscopic length scale $a$. Equations \eqref{eq:cosexp} and \eqref{eq:thetatheta} imply that:
\begin{equation}
\langle \cos(\theta_R - \theta_L) \rangle \sim L^{-1/K} \;.
\end{equation}

Finally, {the RG flow of the weak link term \eqref{eq:Hweaklink} is obtained 
by assuming $1$ as the bare dimension (corresponding to the rescaling of the variable $y$) and $\langle \cos (\theta_R - \theta_L) \rangle$ as the anomalous dimension:}
\begin{equation}\label{eq:KFflow}
\frac{\text{d} J_W}{\text{d} \ell} = \left(1-\frac{1}{K}\right) J_W
\end{equation}
with $\ell=\ln L$. The presence of the weak link term \eqref{eq:Hweaklink} is thus {an irrelevant perturbation for $K<1$: the weak link strength vanishes and cuts the system in two independent parts. 
On the contrary, it is a relevant perturbation for $K>1$: according to this perturbative approach,} the strength of the weak link $J_W$ will be renormalized to larger and larger values.  
To be able to describe the crossover towards transparency, one needs however to go beyond this perturbation on a decoupled system.\\

\paragraph*{(ii)} Step (ii) thus attemps to evaluate \eqref{eq:cosexp} in the presence of $J_W$. We evaluate the propagator at gaussian level, i.e. the model we
consider will be described by: 
\begin{equation}\label{eq:coupJw2D}
 \frac{1}{2} J \int_{L+R} (\boldsymbol{\nabla} \theta)^2 \ud \boldsymbol r + \frac{1}{2} J_W \int_{x=0} (\theta_L(y) - \theta_R(y))^2 \frac{\ud y}{a}  \; .
\end{equation}

Let us first show how to approach the ``pure'' 1d model: 
\begin{equation}\label{eq:pure1d}
  \frac{1}{2} J \int_{L+R} (\nabla \theta)^2 \ud x  \, a + \frac{1}{2} J_W (\theta_L - \theta_R)^2   \; .
\end{equation}
We perform a Hubbard-Stratonovich decoupling of $J_W$ in \eqref{eq:pure1d} introducing a variable $\lambda$:
\begin{equation}
 e^{-\frac{1}{2 T} J_W (\theta_L-\theta_R)^2} \propto \int d \lambda \; e^{-\frac{T}{2 J_W} \lambda^2 + i \lambda (\theta_R-\theta_L)} \; .
\end{equation}
In the following, the will denote $\tilde{J}_W = J_W/T$. 
Then at the action level, we can integrate out the $\theta$-degree of freedom obtaining $\frac{1}{2} \lambda^2 (G_{RR}^0 + G_{LL}^0)$, where $G^0$ is the local $\theta$ propagator with no $J_W$,
from which:
\begin{equation}
 \langle \lambda \lambda \rangle = \left(\frac{1}{\tilde{J}_W} + W \right)^{-1} \text{ with } W= G_{RR}^0+G_{LL}^0 \; . 
\end{equation}
On the other hand, using Dyson equation, one can write $\langle \lambda \lambda \rangle = \langle \lambda \lambda \rangle_0 -  \langle \lambda \lambda \rangle_0^2 G$ 
with $G=\langle (\theta_R - \theta_L)^2 \rangle$. From
\begin{equation}
 \langle \lambda \lambda \rangle = \frac{\tilde{J}_W}{1+\tilde{J}_W W} = \tilde{J}_W - {\tilde{J}_W}^2 G \; ,
\end{equation}
we get:
\begin{equation}\label{eq:G1d}
 G = \frac{1}{\tilde{J}_W} \left(1-\frac{1}{1+\tilde{J}_W W} \right) = \frac{W}{1+\tilde{J}_W W} \;.
\end{equation}
This equation corrects equation \eqref{eq:cosexp}:
\begin{equation} 
\langle \cos(\theta_R - \theta_L) \rangle = e^{-W/2} \rightarrow 
\langle \cos(\theta_R - \theta_L) \rangle =e^{-\frac{W}{2(1+\tilde{J}_W W)}} \; .
\end{equation}
Notice that $\frac{W}{2} = \frac{1}{K} \ln L$, equation \eqref{eq:G1d} is therefore \textit{not} what we want since \eqref{eq:G1d} will always crossover towards $\frac{1}{\tilde{J}_W}$ {irrespectively of $J_W$}.

\begin{figure}
 \centering
  \includegraphics[width=\linewidth]{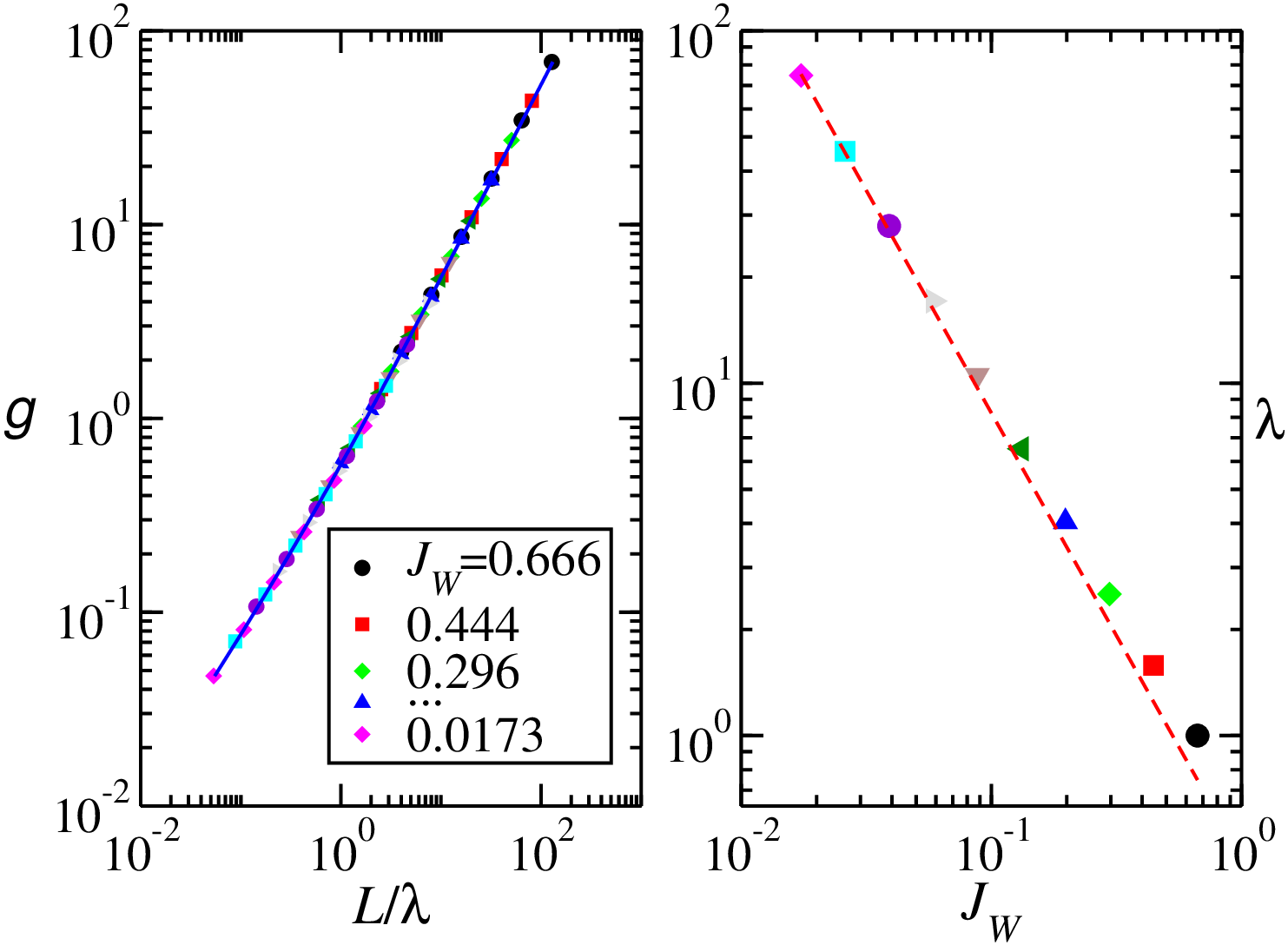}
\caption{(Color online) Scaling behaviour of the correlations accross the weak link, {characterized by the coupling $g=\frac{J_W}{J} L \langle \cos(\theta_R - \theta_L) \rangle$, for different system sizes $L$ from $4$ to $128$ and different values of the weak link strength {$J_W=1.5^{-k}$} from $k=1$ to $k=10$}. Left panel: {When plotted as a function of $L/\lambda$, the data for the coupling $g$ all collapse onto a single scaling curve. The blue line is a fit by the theoretical prediction Eq.~\eqref{eq:correlSC} with {$A_0\approx 0.44$ and $ A_1 \approx 4.1$}.
The crossover lengthscale $\lambda$ is determined {through finite-size scaling and is plotted on the right panel as a function of $J_W$. The red dashed curve shows} the theoretical prediction \eqref{eq:lamWLcorr}.} The 2DXY model with $J=1$ and {$T=0.55$} has been considered{, and the bulk value of $K\approx 4.8$ has been determined by the stiffness along the $y$-axis, $K =\pi \rho_y/T$ at the largest system size $L=128$}.
}
\label{fig:WLcorr}
\end{figure}

The solution is to apply \eqref{eq:pure1d} really in 2d. This amounts to introduce a Hubbard-Stratonovich $\lambda_\omega$ for each Fourier component $\theta_{L,R}(\omega)$
(where $\omega$ is the wave vector associated to the direction $y$). Then, the second term in \eqref{eq:coupJw2D} is replaced by
\begin{equation}
\frac{1}{2 \tilde{J}_W} \sum \lambda_\omega^2 - i \sum \lambda_\omega (\theta_L(\omega) - \theta_R(\omega)) ,
\end{equation}
where we have taken advantage from translation invariance in the $y$-direction. 
The local propagator with no $J_W$ is now replaced by $G^0_{RR}(\omega) = G^0_{LL}(\omega)= \frac{1}{K \vert \omega \vert}$ (i.e. the relevant singular part).
Now \eqref{eq:G1d} changes into 
\begin{equation}
 G(\omega)= \frac{W(\omega)}{1+\tilde{J}_W W(\omega)} = \frac{2}{K} \frac{1}{\vert \omega \vert + 2 \tilde{J}_W /K}\; .
\end{equation}
Then the relevant exponent is:
\begin{equation}\label{eq:expocorr}
 \frac{1}{2} \int_{1/L}^{1} d\omega \, G(\omega) \approx - \frac{1}{K} \ln \left(\frac{1}{L} + \frac{2 \tilde{J}_W }{K}\right) \; ,
\end{equation}
apart from an irrelevant constant.
{Therefore, in the presence of $J_W$, the correlation through the weak link should follow:
\begin{equation}\label{eq:correlsimple}
 \langle \cos(\theta_R - \theta_L) \rangle \propto L^{-1/K} \left( 1+\frac{2}{\pi}\frac{\mathcal J_W}{J} L\right)^{1/K} \;.
\end{equation}}
{In the previous Eq.~\eqref{eq:correlsimple}, $\mathcal J_W$ in the right hand side is the renormalized effective strength of the weak link $\mathcal J_W = J_W \langle \cos(\theta_R - \theta_L) \rangle$. {This approach can be understood as a self-consistent harmonic approximation \cite{PhysRevB.43.344} where, in the calculation of $\langle \cos(\theta_R - \theta_L) \rangle$, the original Hamiltonian \eqref{eq:H2DXY} with a weak link \eqref{eq:Hweaklink} is replaced by an harmonic one \eqref{eq:coupJw2D} with coupling $J_W$ replaced by $\mathcal J_W$.}
We arrive thus at a self-consistent equation for the Kane-Fisher coupling:
\begin{equation}\label{eq:correlSC}
 g = \frac{J_W}{J} L \langle \cos(\theta_R - \theta_L) \rangle = A_0 \left(\frac{L}{\lambda}\right)^{\frac{K-1}{K}} \left(1+ A_1 g\right)^{1/K} \; .
\end{equation}
$A_0$ and $A_1$ are two constants of order one which are difficult to determine theoretically. 
The crossover lengthscale is the so-called ``clutched scale'' of \cite{Yao2016a} and follows:
\begin{equation}\label{eq:lamWLcorr}
 \lambda \sim \left(\frac{J}{J_W}\right)^{K/(K-1)}\;.
\end{equation}
}

We have tested these predictions with Monte Carlo simulations of the 2DXY model with a columnar weak link. The figure \ref{fig:WLcorr} represents the scaling behavior of 
the correlations across the weak link as a function of system size $L$ for different values of the weak link strength $J_W$. When the data for 
{the coupling $g$} are plotted as a function of $L/\lambda$, they all collapse onto a single scaling curve
which agrees very well with the theoretical prediction \eqref{eq:correlSC}, as shown by the blue line. The crossover lengthscale $\lambda$, determined through finite-size scaling, depends only on $J_W$ ($J=1$ has a fixed value) and agrees very well with the theoretical prediction \eqref{eq:lamWLcorr} (red dashed line).

\section{Stiffness}

\begin{figure}
 \centering
  \includegraphics[width=\linewidth]{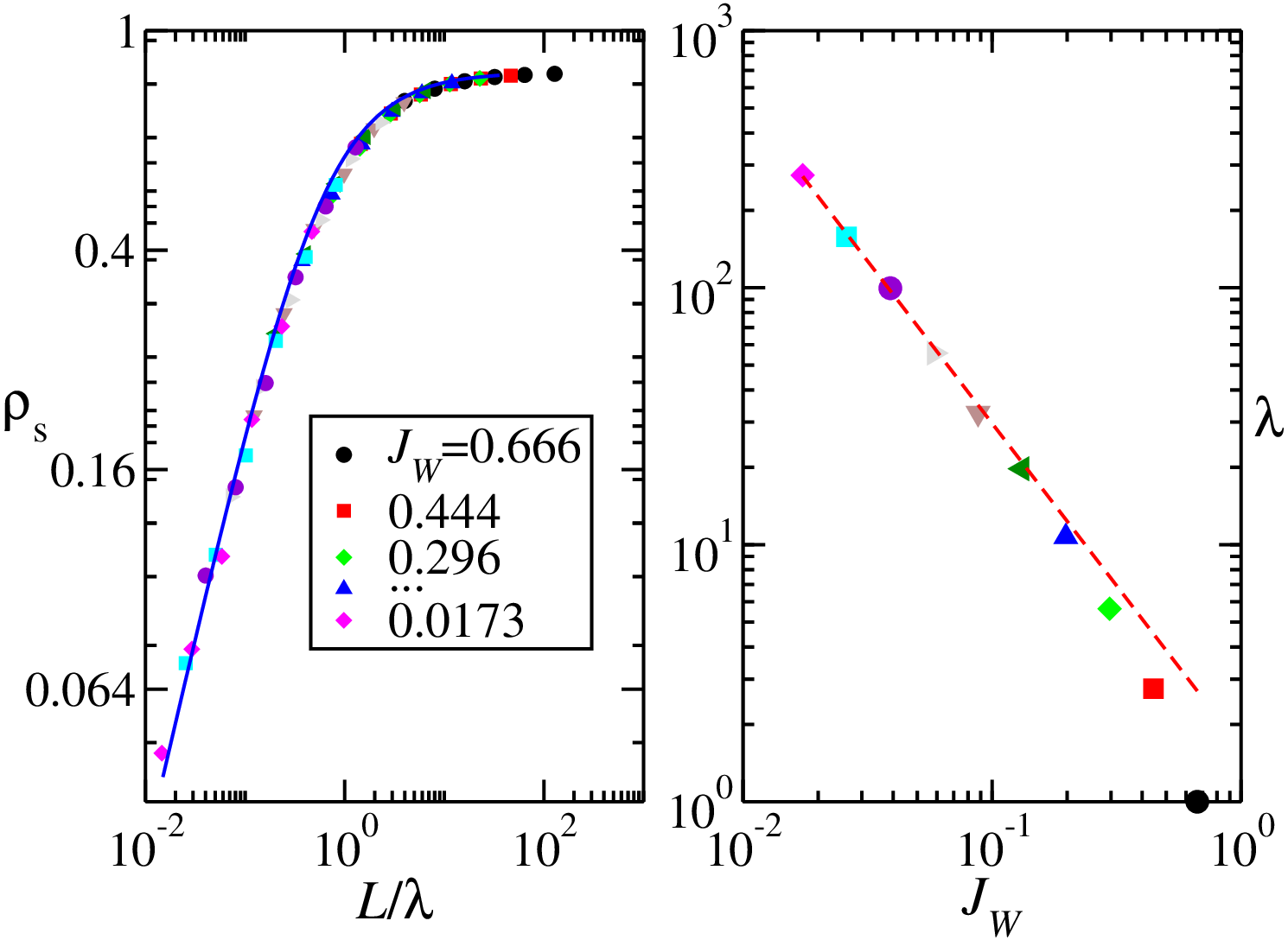}
\caption{(Color online) Scaling behavior of the stiffness $\rho_s$ along the $x$-direction, as a function of system size and for different values of the weak link coupling {$J_W=1.5^{-k}$}, with $k=1$ to $k=10$. {Left panel:} When plotted as a function of $L/\lambda$, the data {for $\rho_s$} collapse onto a single scaling curve well described by the theoretical prediction \eqref{eq:scarhosfunc} {(see the blue line). Right panel: The behavior of the clutch scale $\lambda(J_W)$, determined by finite-size scaling, is well described at small $J_W$ by the theoretical prediction \eqref{eq:lamWLcorr}, as shown by the red dashed line.}
The system size varies from $L=4$ to $L=128$ and the temperature has been fixed to {$T=0.55$} ($J=1$) so that {$K\approx 4.8$}, determined by the stiffness along the $y$-direction {at $L=128$}. The flow of the stiffness is towards transparency, $\rho_s \rightarrow \rho_y$, as expected for $K>1$.}
\label{fig:WLstiff}
\end{figure}

In this section, we want to describe the effect of the columnar weak link on the stiffness along the $x$ axis. Due to the translation invariance along the $y$-direction,
a twist in the $x$-boundary conditions will not induce a current along the $y$ axis, therefore we can consider this as a 1D problem. One can show \cite{refael2013strong} that the stiffness for a 1D chain of size $L$ described by the {harmonic} action:
\begin{equation}\label{eq:harmAction}
 S = \frac{1}{2} \sum_i J_i (\delta \theta_i)^2 \; ,
\end{equation}
with $\delta \theta_i = \theta_{i+1}-\theta_i$, is given by:
\begin{equation}
 \frac{1}{\rho_s} = \frac{1}{L} \sum_i \frac{1}{J_i} \; .
\end{equation}
Therefore, in the case of the 2DXY model with a columnar weak link, we may expect that:
\begin{equation}\label{eq:stiff1D}
 \frac{1}{\rho_s} = \frac{1}{L} \left( \frac{L-1}{J} + \frac{1}{J_W}\right).
\end{equation}
Due to thermal fluctuations, the bulk stiffness is renormalized from $J$ to $\rho_y$. On the other hand, the weak link coupling strength $J_W$ should be replaced by 
\begin{equation}\label{eq:scaJW}
\mathcal J_W = J_W \langle \cos(\theta_R - \theta_L) \rangle 
\end{equation}  
through the Kane-Fisher RG flow {described in the previous section}. 
Incorporating these changes in \eqref{eq:stiff1D} gives:
\begin{equation}\label{eq:scarhosfunc}
 \rho_s = \frac{\rho_y L \mathcal J_W}{L\mathcal J_W + \rho_y} {=  \frac{\rho_y g}{g + \rho_y}}\; .
\end{equation}
This (uncontrolled) approximation can {again} be understood as a self-consistent harmonic approximation \cite{PhysRevB.43.344}, where the original Hamiltonian \eqref{eq:H2DXY} with a weak link \eqref{eq:Hweaklink} is replaced by an harmonic one \eqref{eq:harmAction} with couplings $J$ replaced by $\rho_y$ and $J_W$ replaced by $\mathcal J_W$. On the other hand, the formula \eqref{eq:scarhosfunc} is justified by the fact that, in Kane-Fisher's physics, the weak link does not affect the properties of the bulk, i.e. it does not induce a change of $\rho_y$.

The figure \ref{fig:WLstiff} represents the evolution of the stiffness $\rho_s$ as a function of system size $L$ for different values of the weak link strength $J_W$. When the data are plotted as a function of {$L/\lambda$, with $\lambda(J_W)$ the clutch scale, they collapse onto a single scaling curve which agrees very well with the theoretical formula \eqref{eq:scarhosfunc} with the coupling $g$ given by Eq.~\eqref{eq:correlSC}.} {The clutch scale $\lambda$ depends only on $J_W$ and has been determined through finite-size scaling. At small $J_W$, its behavior agrees well with the theoretical prediction Eq.~\eqref{eq:lamWLcorr}.} In Fig.~\ref{fig:WLstiff}, a flow towards transparency $\rho_s \rightarrow \rho_y$ is clearly observed, as expected for $K>1$.

\section{Power-law weak link: adjustable Kane-Fisher transition}

Up to now, we have been able to investigate only the transparent regime of the Kane-Fisher transition which arises for $K>1$. Indeed, in the 2DXY model, we are constrained
to work at $K>2$, otherwise the quasi long-range correlations are destroyed by the BKT transition \cite{berezinskii1972destruction, kosterlitz1973ordering, kosterlitz1974critical}. Recently, Prokof'ev, Svistunov and colleagues proposed \cite{Yao2016a}, in the context of the superfluid-insulator transition in 1D quantum disordered bosons, that a weak link whose strength decreases algebraically with system size $J_W=J_0 L^{-\alpha}$, $\alpha>0$, induces a Kane-Fisher transition \cite{Kane1992a, kane1992transmission} at a threshold $K_c=\frac{1}{1-\alpha}>1$. In this section, we address this problem on the basis of our previous theoretical arguments and we show that it allows us to observe and characterize the Kane-Fisher transition in the 2DXY model.

\subsection{Correlations across a power-law weak link}

\begin{figure}
 \centering
  \includegraphics[width=\linewidth]{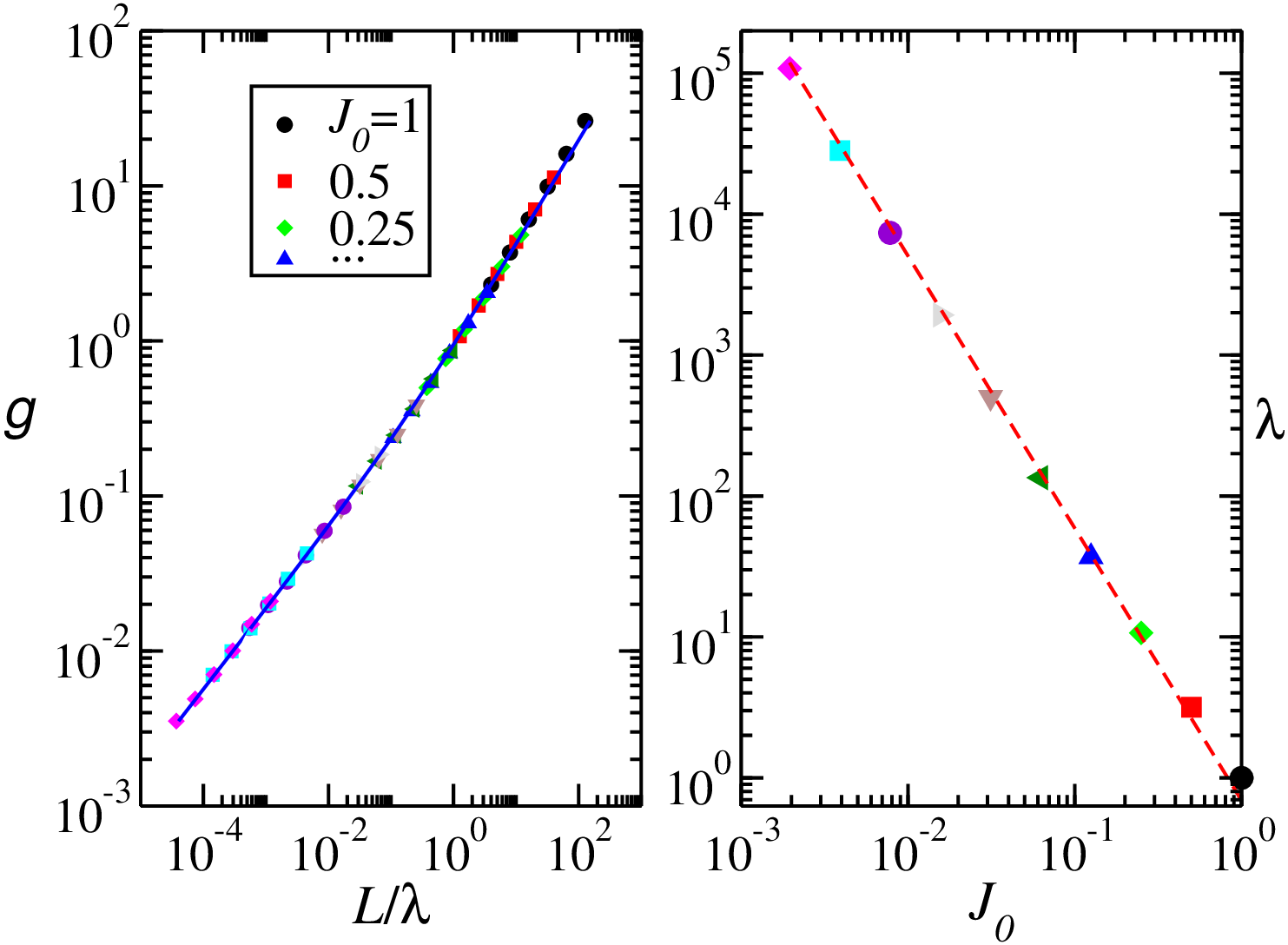}
\caption{(Color online) Scaling behavior of the correlations across a power-law weak link $J_W=J_0 L^{-\alpha}$ as a function of system size and for different values of $J_0=2^{-k}$, from $k=0$ to $k=9$. {Left panel: When the coupling {$g= \frac{J_0}{J} L^{1-\alpha} \langle \cos(\theta_R - \theta_L) \rangle$} is plotted as a function of $L/\lambda$ with $\lambda$ the clutch scale which depends only on $J_0$, the data collapse onto a single scaling curve well fitted by the equation \eqref{eq:scagalpha} with $A_0\approx 0.66$ and $A_1\approx3.9$, as shown by the blue line. Right panel: The behavior of the clutch scale $\lambda$, determined through finite-size scaling, as a function of $J_0$ is well fitted by Eq.~\eqref{eq:lambdaalpha}, as shown by the red dashed line.} The system size varies from $L=4$ to $L=128$. $\alpha=0.25$, $T=0.6$ and $J=1$ so that $K=4.3$, determined through the stiffness along the $y$-axis.}
\label{fig:WLcorr-alpha}
\end{figure}

\begin{figure*}
 \centering
  \includegraphics[width=0.32\linewidth]{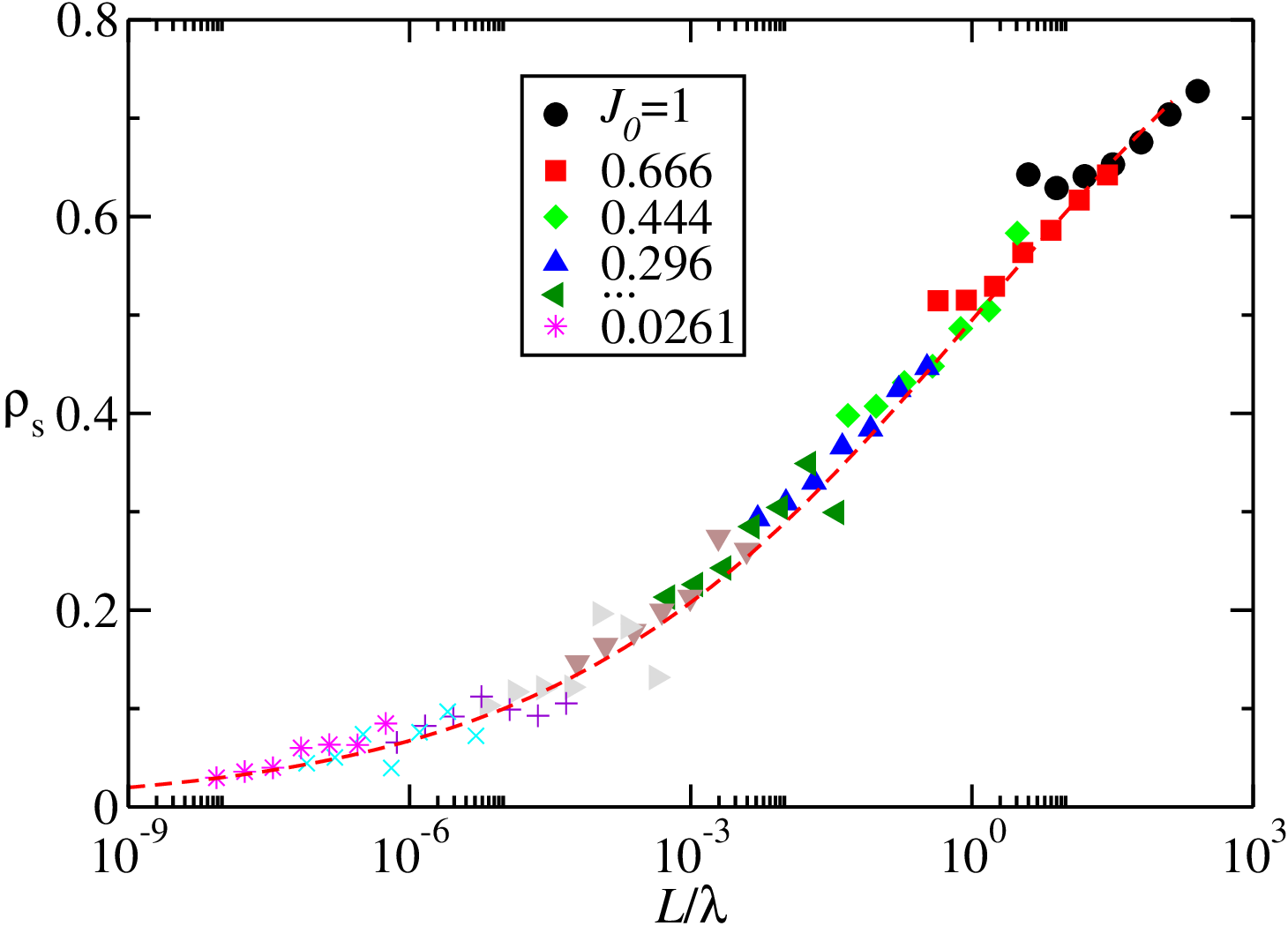} 
    \includegraphics[width=0.32\linewidth]{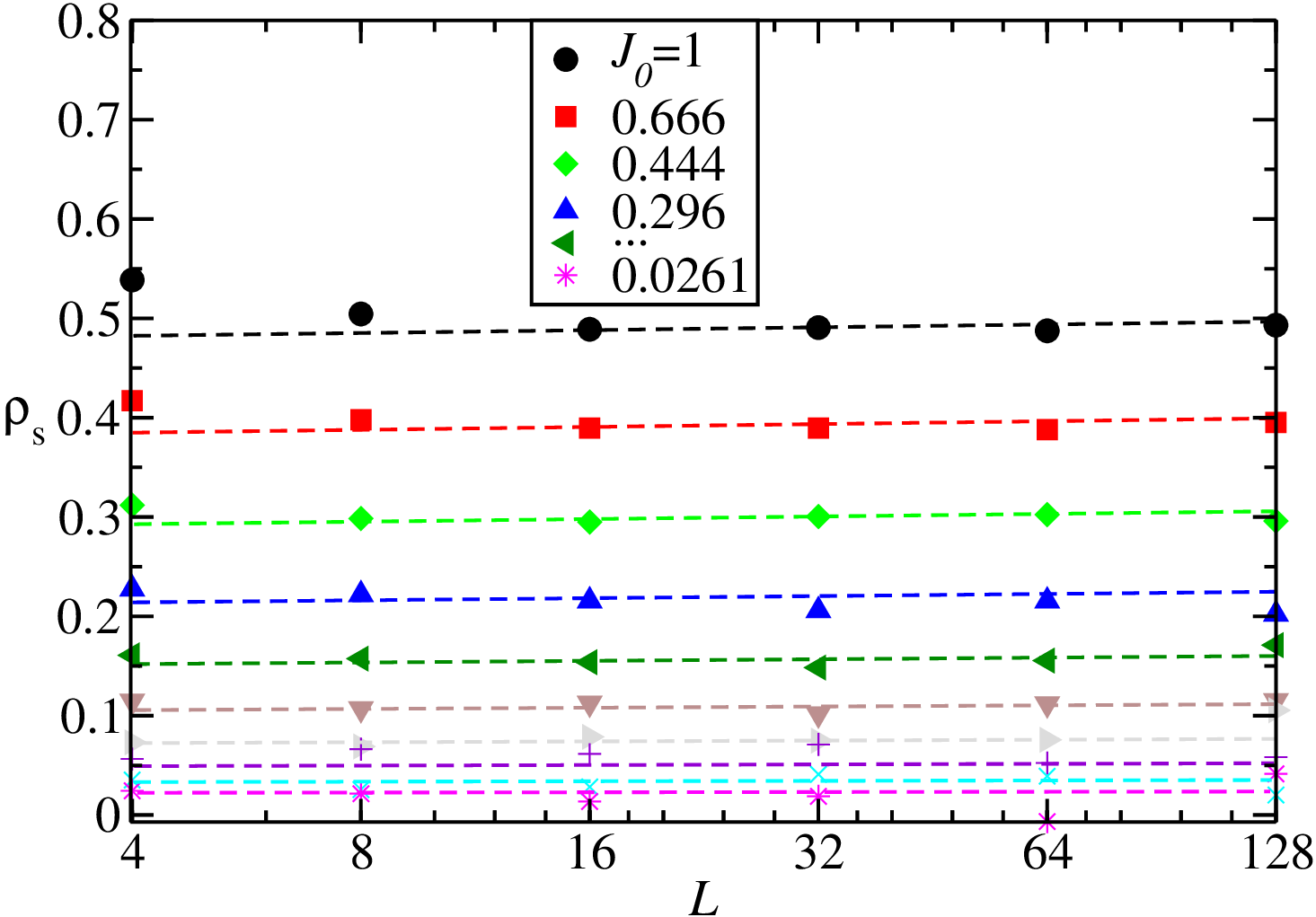}
  \includegraphics[width=0.32\linewidth]{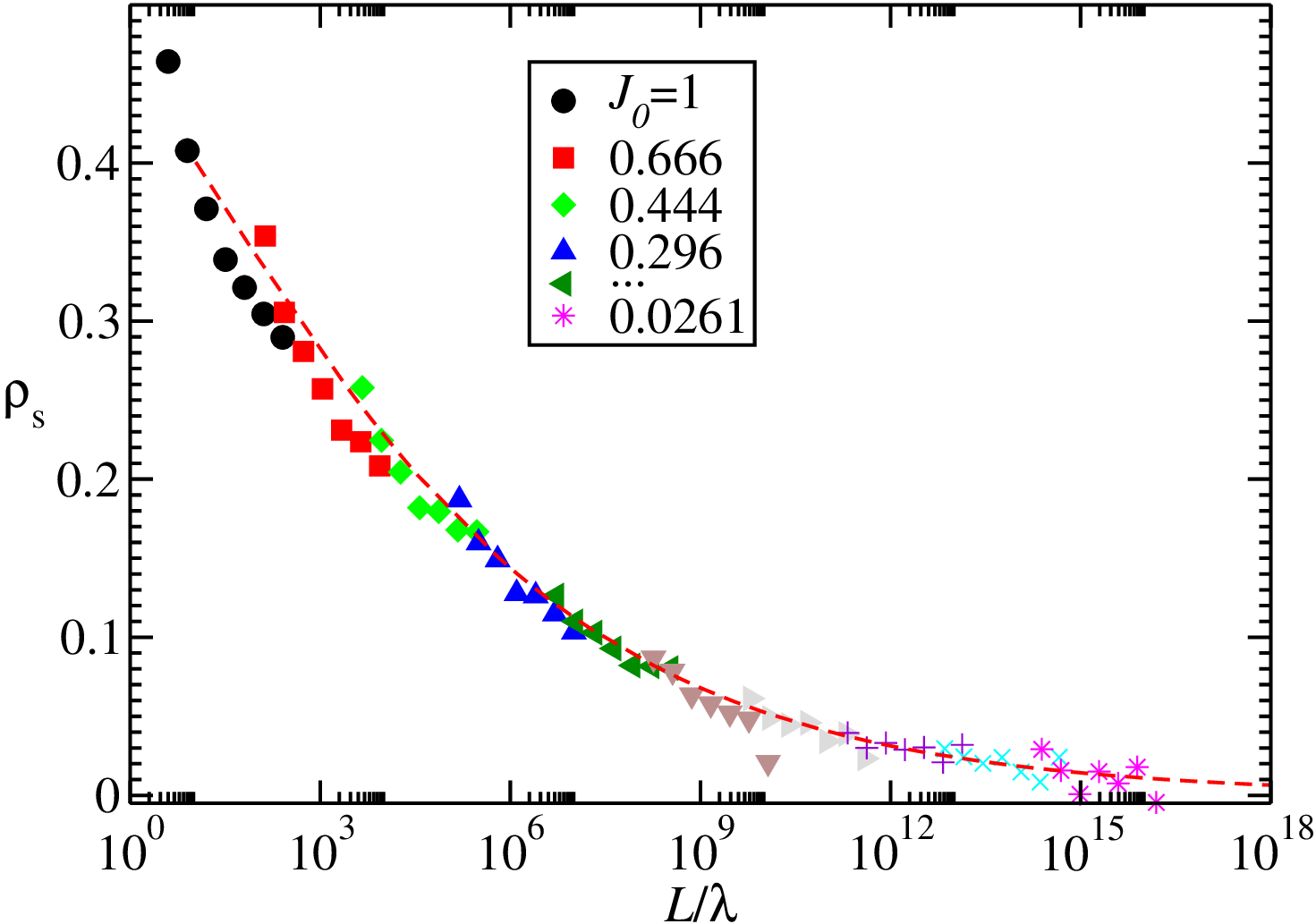}
\caption{(Color online) Kane-Fisher transition on the 2DXY model with a power-law weak link $J_W= J_0 L^{-\alpha}$. $\alpha=0.75$ implies a critical $K_c=\frac{1}{1-\alpha} = 4>2$ so that the Kane-Fisher transition from a flow towards transparency to a flow towards a cut can be observed in the quasi-ordered phase of the BKT transition $K>2$. The left panel shows the data for $T=0.2$ where {$K\approx 14.9>K_c$}. When plotted as a function of $L/\lambda$ with $\lambda$ the clutch scale given by eq.~\eqref{eq:lambdaalpha}, the data collapse onto the scaling function \eqref{eq:scarhosfunc} shown by the red dashed line. The right panel shows the case of {$K \approx 2.8<K_c$} ($T=0.8$) where the flow is towards a cut $\rho_s \rightarrow 0$ when $L\gg \lambda$, and follows the theoretical prediction \eqref{eq:scarhosfunc} (red dashed line). The middle panel corresponds to the vicinity of the Kane-Fisher transition, {$K\approx 4.3\approx K_c$} ($T=0.6$) for which Kane-Fisher renormalization of the weak link is almost irrelevant and the stiffness depends only weakly on system size and follows again \eqref{eq:scarhosfunc} (see the dashed lines). The values of {$J_0=1.5^{-k}$} with $k=0$ to $k=9$, and the system size varies from $L=4$ to $L=256$ in the right and left panels and up to $L=128$ in the middle panel.}
\label{fig:stiff-alpha}
\end{figure*}

Let us now consider a power law weak link $J_W=J_0 L^{-\alpha}$ and see how the theoretical predictions of section \ref{sec:constWLcorr} are modified.
Inserting $J_W = J_0 L^{-\alpha}$ in the equation \eqref{eq:correlsimple} for the weak link correlations, we get a new self-consistent equation for the coupling $g = \frac{J_0}{J} L^{1-\alpha} \langle \cos(\theta_R - \theta_L) \rangle$ :
\begin{equation}\label{eq:scagalpha}
 g=  A_0 \left(\frac{L}{\lambda}\right)^{\frac{K(1-\alpha)-1}{K}} \left(1+ A_1 g\right)^{1/K} \; ,
\end{equation}
with the clutch scale 
\begin{equation}\label{eq:lambdaalpha}
 \lambda \sim \left(\frac{J}{J_0}\right)^{\frac{K}{K(1-\alpha)-1}}\;.
\end{equation}

The figure \ref{fig:WLcorr-alpha} represents this scaling behavior for $\alpha=0.25$ and $T=0.4$. When plotted as a function of $L/\lambda$, with $\lambda$ the clutch scale given by \eqref{eq:lambdaalpha}, the data for the coupling $g$ all collapse onto a single scaling curve which agrees well with the theoretical prediction \eqref{eq:scagalpha}, as shown by the blue line.

\subsection{Kane-Fisher transition on the stiffness}

In this section, we show how a power-law weak link $J_W=J_0 L^{-\alpha}$ allows for the observation of the Kane-Fisher transition in the 2DXY model. {According to equations \eqref{eq:scarhosfunc} and \eqref{eq:scagalpha} the evolution of the stiffness as a function of system size depends on the variable $L/\lambda$ with the clutch scale given by Eq.~\eqref{eq:lambdaalpha}. This implies a Kane-Fisher transition at
\begin{equation}\label{eq:Kcalpha}
 K_c= \frac{1}{1-\alpha} \; .
\end{equation}
If $K>K_c$, the flow of the stiffness is towards transparency, while for $K<K_c$, the flow is towards a cut.} 

Setting $\alpha>0.5$, we should be able to observe the Kane-Fisher transition in the 2DXY since $K_c(\alpha)>2$ is larger than the threshold of the BKT transition. The figure \ref{fig:stiff-alpha} shows the results in the case $\alpha=0.75$ where $K_c(\alpha)=4$ (see Eq.~\eqref{eq:Kcalpha}). For $T=0.2$, {$K\approx 14.9>K_c$}, we observe clearly that the stiffness converges towards its transparent value $\rho_s \rightarrow \rho_y$ as a scaling function of the variable {$L/\lambda$}. The agreement with the theoretical prediction {\eqref{eq:scarhosfunc}} shown by the red dashed curve is excellent. On the other hand, for $T=0.8$, {$K\approx 2.8<K_c$}, $K(1-\alpha)-1<0$ and the flow is towards a cut. The data, when plotted as a function of {$L/\lambda$ with $\lambda$ given by Eq.~\eqref{eq:lambdaalpha} all collapse onto a single scaling curve given again by \eqref{eq:scarhosfunc}}. This implies that the stiffness vanishes as a power law with system size $\rho_s \approx J_0 L^{\frac{K(1-\alpha)-1}{K}}$ at large $L$.

Close to the threshold {$K\approx 4.3 \approx K_c$} for $T=0.6$, Kane-Fisher renormalization of the weak link is {almost} absent, and the stiffness is a non trivial number between $0$ and $\rho_y$ given by Eq.~\eqref{eq:scarhosfunc},
a prediction which agrees well with the numerical data (see the dashed lines in the middle panel).

\section{Vortex energy argument}
 
This final section aims at giving a thermodynamical argument for the adjustable $K_c=1/(1-\alpha)$ of the Kane-Fisher transition in the case of a power-law weak link.   
{It is well-known that the BKT transition is driven by topological vortex excitations  \cite{berezinskii1972destruction, kosterlitz1973ordering, kosterlitz1974critical}. The original argument for the BKT transition \cite{kosterlitz1973ordering} compares the energy cost of a single vortex excitation with its entropy, which are both found to scale logarithmically with system size in two dimension, so that the free energy reads $ F = E - TS = (\pi J - 2 T) \ln L$. For $K=\pi J/T>2$, we have a proliferation of single vortices and the quasi long-range order is destroyed. A similar argument can be made for the Kane-Fisher transition in 1D quantum systems \cite{PhysRevB.53.13091}, where vortices in the $x,\tau$ plane ($\tau$ being the imaginary time) are then constrained to locate only in the vicinity of the columnar weak link. This constraint changes the entropy per vortex to $S =\ln L$ since there are only $L$ different configurations of the vortex, instead of $L^2$. We thus recover the threshold $K_c=1$ for the standard Kane-Fisher transition. It is however not clear how to extend these ideas to the case of a power-law weak link. As we will show, the energy of a single vortex in the case of a power-law weak link depends in a non-trivial manner on $\alpha$ and $L$ and this allows us to recover $K_c= 1/(1-\alpha)$ for the threshold of the Kane-Fisher transition in this case.}

{The issue is to} evaluate the energy of a single vortex in a configuration which consists in a slice $B$ of width $2d$ with coupling $J_W$ {between two $L^2$ systems $A$ and $A'$} with coupling $J$, with $J_W/J\equiv W \ll 1$. {For simplicity}, the vortex is supposed to be located in the middle of $B$. 
We first use the standard analogy with an electrostatic problem (see for example \cite{benfatto2013berezinskii}). The vortex is characterized by the circuitation:
\begin{equation}
 \oint_\mathcal C  \boldsymbol{\nabla} \theta \cdot \text{d}\boldsymbol{l}= \int_\mathcal S (\boldsymbol{\nabla} \times \boldsymbol{j}_\perp) =  2 \pi \; ,
\end{equation}
where the current field 
$\boldsymbol{j}_\perp =\boldsymbol{\nabla} \theta  $. We introduce the scalar function $\Phi$ such that $\boldsymbol{j}_\perp = \boldsymbol{\nabla} \times (\hat{z} \Phi)= (\partial_y \Phi, -\partial_x \Phi, 0)$. Therefore, $\boldsymbol{\nabla} \times  \boldsymbol{j}_\perp = (0,0,-\nabla^2 \Phi)$, i.e. $\Phi$ satisfies the Poisson equation:
\begin{equation}
 \nabla^2 \Phi = -2\pi \delta(\boldsymbol r) \; .
\end{equation}
In the following, we denote by {$\boldsymbol D= -\boldsymbol \nabla \Phi$.}
The conditions at the boundary are:{
\begin{eqnarray}\label{eq:boundcondi}
 J D_y^A  &=& J_W D_y^B \; ,\\
  D_x^A &=& D_x^B \; ,
\end{eqnarray}}
which express the current conservation $(J_{x,y} \nabla_{x,y} \theta)_A = (J_{x,y} \nabla_{x,y} \theta)_B$ with $(J_x)_A=J$ and $(J_x)_B=J_W$, while $J_y=J$ everywhere.
Thus, this problem is equivalent to a dielectric problem with {$\boldsymbol D$} interpreted as the electric displacement field, $1/J(\boldsymbol r)$ the analog of the permittivity and {$\boldsymbol E=J \boldsymbol D$} the analog of the electric field.

In the appendix \ref{sec:elecfield}, we derive the explicit form of the electric {displacement} field through the method of image charges.
The energy is then evaluated as 
\begin{equation}
E_\text{vort}  = \frac{1}{2} \int_{A+B}J(\boldsymbol r) {\boldsymbol D}(\boldsymbol r)^2 d\boldsymbol r \approx  \frac{1}{2} \int_{A}J(\boldsymbol r) {\boldsymbol{D}}(\boldsymbol r)^2 d\boldsymbol r .
\end{equation}

The result is that for a constant weak link $J_W$ 
\begin{equation}
 E_\text{vort} \approx J \ln L \; .
\end{equation}
The entropy of such a vortex constrained on a slice B is $\ln L$, therefore the Kane-Fisher transition appends when the energy and entropy terms compensate exactly, i.e. at $ \pi J = T$, or $ K_c=1$. On the other hand, for $J_W = J_0 L^{-\alpha}$, we find (see appendix \ref{sec:elecfield}):
\begin{equation}
 E_\text{vort} \approx J \;  \ln \left( L^{1-\alpha} \right) \; ,
\end{equation}
leading to $K_c= \frac{1}{1-\alpha} $ in the case of a power-law weak link.

\section{Conclusion}

In this paper, we have studied a simple model underlying the scratched-XY scenario \cite{Yao2016a} for the strong disorder regime of the 1D superfluid-insulator transition \cite{Giamarchi1987a, Giamarchi1988a, ristivojevic2012phase, Altman2004a, Altman2008a, Altman2010a, refael2013strong, Yao2016a, pielawa2013numerical, PhysRevLett.109.265303, gerster2016superfluid, doggen2017weak, pfeffer2018strong}. The model consists of a weak link whose strength decreases algebraically with the system size $ J_W \sim L ^ {-\alpha} $, in an otherwise \textit{clean} system. Using the analogy between 1D quantum systems and the classical 2DXY model, where the weak link is replaced by a weak link column, we were able to describe a Kane-Fisher transition \cite{Kane1992a, kane1992transmission} from a transparent regime for $K>K_c$ to a perfect cut for $K<K_c$, with an adjustable $K_c=1/(1-\alpha)$ depending on $\alpha$. Our theory is found in very good agreement with the results of Monte Carlo numerical simulations and accounts for the full crossover from weak link physics to transparency.

This work clarifies two important assumptions at the basis of the scratched-XY {scenario} \cite{Yao2016a}. First, the ``clutch scale'', describing the crossover of the superfluid density, is given by equation \eqref{eq:lambdaalpha}, and second the validity of the ``classical flow'', i.e. formula \eqref{eq:scarhosfunc}, has been checked with numerical data (see figures \ref{fig:WLstiff} and \ref{fig:stiff-alpha}). Importantly, the coupling $g = \frac{J_0}{J} L^{1-\alpha} \langle \cos(\theta_R - \theta_L) \rangle$ is the analog of the variable $1/w$, Eq.~(2.13) of \cite{Yao2016a}. From our Eq.~\eqref{eq:scagalpha}, the logarithmic derivative of the coupling $g$ with respect to $L$ follows:
\begin{equation}\label{eq:RGg}
 \frac{\partial g}{\partial \ln L}=\frac{K(1-\alpha)-1}{K- g A1/(1+A1g)} g \; ,
\end{equation}
which corrects the renormalization flow for $w$, Eq.~(2.21c) of \cite{Yao2016a}, with $\zeta=1-\alpha$. Notice
that in our Eq.~\eqref{eq:RGg}, the denominator varies from $K$ at small $g$ (i.e. $L\ll \lambda$) to $K-1$ at large $g$ ($L\gg \lambda$), contrary to Eq.~(2.21c) of \cite{Yao2016a} where it is always $K-1$.

{While our results validate several predictions made in \cite{Yao2016a}, the Kane-Fisher transition that we find is clearly distinct from a BKT transition such as the 1D superfluid-insulator transition. In particular, the stiffness in the cut regime $K<K_c$ decreases as a power law with system size instead of the exponential decay characteristic of the insulating phase. Moreover, at the transition, we do not observe the strong (logarithmic) finite-size effects expected for a BKT transition, but the stiffness stays constant as a function of system size and depends crucially on the microscopic strength of the weak link (see the middle panel of figure \ref{fig:stiff-alpha}).  

{In fact}, the  scratched-XY model \cite{Yao2016a} incorporates another important ingredient: the bulk of the system should not be considered clean, but instead incorporates the effect of many weak links. In \cite{Yao2016a}, the authors propose that such a bulk can be described by a clean bulk with a renormalized coupling $J(L)$ accounting for the other weak links self-consistently. We would like to stress that this is an uncontrolled assumption which has not been tested yet. In \cite{vasseur2017healing}, the effect of a weak link in a disordered XXZ chain was studied. It was argued, in the presence of bond-disorder, that a weak link is healed even in the antiferromagnetic case (where $K<1$), in contrast to the clean bulk case where healing occurs only in the ferromagnetic case ($K>1$). Moreover, the corresponding clutch scale has a logarithmic dependency on the weak link strength which is very different from the algebraic dependence found here \eqref{eq:lambdaalpha}. More work is therefore needed to describe the Kane-Fisher physics in the presence of a disordered bulk. }

{As a final remark, let us discuss the recent numerical study \cite{pfeffer2018strong} of the scratched-XY model with power-law distributed weak links. 
In the regime where the arguments of \cite{Yao2016a} predict a transition, the numerical results of \cite{pfeffer2018strong} show very strong finite size effects which practically prevent to distinguish the insulating behavior from the superfluid one with the available systems sizes (as large as $L=512$). It is well known that BKT transitions have strong logarithmic finite-size corrections at criticality, and their precise knowledge is important to characterize numerically the critical behavior (see e.g. \cite{hsieh2013finite}). In \cite{Yao2016a}, the authors have made such a prediction, however based on the assumptions already discussed concerning the clutch scale, the classical flow and the self-consistent bulk. Our theory has clarified the first two assumptions
and in particular corrects the renormalization flow of $g$, Eq.~\eqref{eq:RGg}, which may change the logarithmic corrections at criticality. It would be interesting to extend our approach to the case of a power-law weak link in a disordered bulk, in particular to assess the relevance of the interplay between different weak links in providing the insulating behavior.}

\begin{acknowledgments}
We thank F. Alet, L. Benfatto, S. Capponi, E.V.H. Doggen and N. Laflorencie for  discussions. G.L. acknowledges
an invited professorship at Sapienza University of Rome. We
thank CALMIP for providing computational resources. This work is supported by the
French ANR program COCOA (Grant No. ANR-17-CE30-0024-01) and  
by  Programme  Investissements  d'Avenir  under
Program No. ANR-11-IDEX-0002-02, Reference No. ANR-10-LABX-0037-NEXT. 
\end{acknowledgments}

\appendix

\section{Energy of a vortex}\label{sec:elecfield}

\subsection{Electric {displacement} field through the method of image charges}

In this appendix, we describe first how to find the {$D$}-field defined by $$\boldsymbol \nabla {\boldsymbol D} = 2 \pi \delta(\boldsymbol r) $$ with the boundary conditions described by \eqref{eq:boundcondi}.
The simplest way is to use the method of image charges and to put the origin $(0,0)$ at the right interface (the vortex is in $(-d,0)$). $B$ is described by charges $\alpha_n$ 
with $\alpha_0=1$ in $(-d,0)$, $n>0$ in $\boldsymbol{r}_n=(-d-2nd,0)$ and $n<0$ in $\boldsymbol{r}_{n}=(-d+2\vert n \vert d,0)$. The symmetry implies $\alpha_n = \alpha_{-n}$ (it is the complication as compared to the 
case of a single interface). $A$ (i.e. the domain $x<-2d$) is described by {charges $\beta_n$, $n\le 0$, in $\boldsymbol{r}_{n}$, and $A'$ ($x>0$) by charges $\beta_n$, $n\ge 0$ in $\boldsymbol{r}_n$}. Note that $\alpha_n$ and $\alpha_{-n-1}$ (and $\beta_n$) have singularities $[(x-d-2nd)^2+y^2]^{-1}$ and 
$[(x-d+(2\vert n\vert +2)d)^2+y^2]^{-1}$ which have the same dependence in y at the interface $x=0$ (at the right interface: $[d^2(1+2nd)^2+y^2]^{-1}$ and $[(1+2\vert n\vert)^2 d^2+y^2]^{-1} $).
There are two equations to consider: 
\begin{eqnarray}
 W {D_y}^B&=&{D_y}^A  \\
 {D_x}^B&=&{D_x}^A \; ,
\end{eqnarray}
with $W=J_W/J$ which involve ($\alpha_0$, $\alpha_{-1}$, $\beta_0$), ... , ($\alpha_n$, $\alpha_{-n-1}$, $\beta_n$). Using ${D_x} = y/[(x-r_n)^2 + y^2]$ and ${D_y} = (x-r_n)/[(x-r_n)^2 + y^2]$, we get:
\begin{eqnarray}
 W(\alpha_n y + \alpha_{-n-1} y ) & = & \beta_n y  \\
 \alpha_n r_n + \alpha_{-n-1} r_{-n-1} &=& \beta_n r_n  .
\end{eqnarray}
At this point, $d$ has disappeared (it will appear in the UV cutoff $d\ge a$).
\begin{eqnarray}
 \alpha_n + \alpha_{-n-1}  & = & W^{-1} \beta_n   \\
 - \alpha_n (1+2n) + (1+2n) \alpha_{-n-1} &=& -\beta_n (1+2n)  , 
\end{eqnarray}
thus:
\begin{eqnarray}
 \alpha_n + \alpha_{-n-1}  & = & W^{-1} \beta_n   \\
 \alpha_n - \alpha_{-n-1} &=& \beta_n  .
\end{eqnarray}
The solutions are:
\begin{eqnarray}
 \alpha_n  & = & \left(W^{-1}+1\right) \frac{\beta_n}{2}   \\
 \alpha_{-n-1} &=& \left(W^{-1}-1\right) \frac{\beta_n}{2}  ,
\end{eqnarray}
which we rewrite as:
\begin{eqnarray}
 \beta_n  & = & 2 \alpha_n  \frac{W}{1+W}   \\
 (\alpha_{n+1}\equiv) \quad \alpha_{-n-1} &=& \alpha_n \frac{1-W}{1+W}  .
\end{eqnarray}
With $\alpha_0=1$,
\begin{eqnarray}
\alpha_{n} &=& \left(\frac{1-W}{1+W}\right)^n  \\
 \beta_n  & = & 2 \frac{W}{1+W} \left(\frac{1-W}{1+W}\right)^n   .
\end{eqnarray}
\begin{itemize}
\item For $W=1$ (i.e. $J_W=J$) we have $\alpha_0=\beta_0=1$ and $\alpha_n=\beta_n=0$ for $n\ge1$.
\item $W=0$, $\alpha_n=1$ and $\beta_n=0$ (the field is entirely confined in B).
\item $W\ll 1$, $\alpha_n \approx (1-2W)^n$ and $\beta_n = 2 W (1-W)(1-2W)^n$.
\end{itemize}
Finally the field in $A$ is:
\begin{eqnarray}\label{eq:EinA}
  {\boldsymbol D_A}(\boldsymbol r ) &=& \sum_{n\ge0} \beta_n \frac{\boldsymbol r - \boldsymbol r_n}{\vert \boldsymbol r - \boldsymbol r_n\vert^2} \nonumber \\
  &=& 
  \sum_{n \ge 0} \beta_n \frac{[x+d(2n+1),y]}{(x+d(2n+1))^2+y^2},
\end{eqnarray}
with 
\begin{equation}
\beta_n = \frac{2 W}{1+W} \left(\frac{1-W}{1+W}\right)^n.
\end{equation}

\subsection{Electrostatic potential energy}

The second step is now to evaluate {$$\frac{1}{2} J \int_A d\boldsymbol r \; \boldsymbol{D}_A^2 (\boldsymbol r).$$}

\subsubsection{Singular and regular terms}
A direct evaluation of the sum involved in {$\boldsymbol D_A$} is difficult. Instead, one approximation that we can make is the following:
\begin{equation}
{\boldsymbol D_A} \approx \sum_{n< n_c} \beta_n \frac{\boldsymbol r}{r^2} + \sum_{n\ge n_c} \beta_n \frac{-\boldsymbol r_n}{{r_n}^2} ,
\end{equation}
where  $n_c=r/(2d)$ such that $n\ll n_c \Leftrightarrow \vert \boldsymbol r_n \vert \ll  \vert \boldsymbol r \vert $. The idea is that 
\begin{eqnarray}
{\boldsymbol D_\mathrm{sing}} & \equiv &\sum_{n< r/(2d)} \beta_n \frac{\boldsymbol r}{r^2} \nonumber \\
&=& 
\frac{2W}{1+W} \sum_{n< r/(2d)} p^n \frac{\boldsymbol r}{r^2} \nonumber \\
&=& \left(1-p^{r/(2d)}\right) \frac{\boldsymbol r}{r^2}\; ,
\end{eqnarray}
is the important term with respect to 
${D_\mathrm{reg}} \hat{x}/d $
with 
\begin{equation}
{D_\mathrm{reg}} \equiv \frac{2 W}{1+W} \sum_{n\ge r/(2d)} p^n \frac{1}{2n+1}.
\end{equation}
{In the previous equations, $p=(1-W)/(1+W)$}. 

\subsubsection{Regular term}
The ${D}_\mathrm{reg}$ term can be evaluated as follows:
\begin{eqnarray}
{D}_\mathrm{reg} &=&  \frac{2 W}{1+W} \frac{1}{\sqrt{p}} \sum_{n\ge r/(2d)} \frac{\sqrt{p}^{2n+1}}{2n+1}  \nonumber \\
&=&\frac{W}{1+W} \frac{1}{\sqrt{p}} \sum_{n> r/(2d)} \frac{p^{n}}{n} \; . 
\end{eqnarray}
Let's denote $I_n\equiv \sum_{k\ge0}^n p^k$. It is clear that 
\begin{equation}
J_n=\sum_{k\ge0}^n \frac{p^{k+1}}{k+1} = \int_0^p I_n \; dp'.
\end{equation}
Since $I_n = \frac{1-p^{n+1}}{1-p} $, 
\begin{equation}
J_n= - \ln (1-p)-\int_0^p \frac{{p'}^{n+1}}{1-p'}dp' ,
\end{equation}
where the first term in the right hand side is $J_\infty$. 
\begin{equation}
\sum_{n> r/(2d)} \frac{\sqrt{p}^{n}}{n} = J_\infty - J_{r/(2d)-1}=\int_0^p \frac{{p'}^{r/(2d)}}{1-p'}dp'.
\end{equation}
Using $p\approx 1-2 W \approx 1- 2 J_0 L^{-\alpha}$, we can rewrite the previous integral as:
\begin{equation}
\int_0^p \frac{{p'}^{r/(2d)}}{1-p'}dp' = \int_{2 J_0 L^{-\alpha}}^1 \frac{(1-u)^{r/(2d)}}{u} du.
\end{equation}
$(1-u)^{r/(2d)} \approx 0$ for $u\gg 2d/r$, while $\approx 1$ for $u \ll 2d/r$, therefore we can approximate the last integral as:
\begin{eqnarray}
\int_{2 J_0 L^{-\alpha}}^1 \frac{(1-u)^{r/(2d)}}{u} du &\approx& \int_{2 J_0 L^{-\alpha}}^{2d/r} \frac{1}{u} du \nonumber \\
&=& \ln \left( \frac{ L^\alpha d}{r J_0} \right)  .
\end{eqnarray}
Finally, 
\begin{equation}
{D}_\mathrm{reg} \approx  W  \ln \left( \frac{ L^\alpha d}{r J_0} \right) \approx J_0 L^{-\alpha} \ln  \left( \frac{ L^\alpha d}{r J_0} \right) .
         \end{equation}

\subsubsection{Vortex energy}

The vortex energy is therefore given by the singular part:
\begin{eqnarray}\label{eq:intEsing}
E_\text{vort} &\approx& \frac{1}{2} J \int_A d\boldsymbol r \; {\boldsymbol{D}}_\mathrm{sing}^2 (\boldsymbol r)  \nonumber \\
&\approx& \int_d^L \left(1-p^{r/(2d)}\right) \frac{1}{r} dr \; .
\end{eqnarray}
We have 
\begin{eqnarray}
p^{r/(2d)} &\approx& (1-2W)^{r/(2d)} \nonumber \\
&\approx& (1-2 J_0 L^{-\alpha})^{r/(2d)} \nonumber \\
&\approx& e^{-r J_0 /(d L^\alpha)}.
\end{eqnarray}
Therefore, 
\begin{eqnarray}
\text{for } r &\gg&  L^\alpha d/J_0\text{, }\; p^{r/(2d)} \approx 0 \, ,\\
\text{while for }  r &\ll& L^\alpha d/J_0\text{, } \;p^{r/(2d)} \approx 1.
\end{eqnarray}
Hence, the integral \eqref{eq:intEsing} can be approximated by:
\begin{eqnarray}
 E_\text{vort} &\approx& 
\int_{L^\alpha d/J_0}^L \frac{1}{r} dr \nonumber \\
&=& \ln \left(\frac{L J_0 }{L^\alpha d}\right) \nonumber \\
&=& (1-\alpha) \ln L +\mathrm{cste}.
\end{eqnarray}

\end {document}